\documentclass[%
 reprint,
 superscriptaddress,
 amsmath,amssymb,
prl,
longbibliography
]{revtex4-1}

\usepackage[colorlinks=true,linkcolor=blue,citecolor=red]{hyperref}
\usepackage{bm}

\usepackage[english]{babel}
\usepackage{microtype}
\usepackage{indentfirst}
\usepackage{graphicx}
\usepackage{placeins}
\usepackage{siunitx, booktabs}
\usepackage{tabularx}
\usepackage[caption=false]{subfig}
\usepackage{amsmath}
\usepackage{array}
\usepackage{dcolumn}
\usepackage{bm}
\usepackage{setspace}
\usepackage{amssymb}
\usepackage{braket}
\usepackage{xcolor}
\usepackage{color,soul}
\usepackage{placeins}
\usepackage{eqnarray,amsmath}
\newcommand{\etal}{{\it et al.}}

\newcommand{\FIG}[1]{\textcolor{red}{Fig. }}
\newcommand{\TAB}[1]{\textcolor{violet}{Table }}

\begin{document}

\title{Out-of-plane polarization and topological magnetic vortices in multiferroic CrPSe$_3$}

\author{Weiwei Gao}
\address{Key Laboratory of Materials Modification by Laser, Ion and Electron Beams, Ministry of Education, Dalian University of Technology, Dalian 116024, China}
\address{Center for Computational Materials, Oden Institute for Computational Engineering and Sciences, The University of Texas at Austin, Austin, TX 78712}

\author{Jijun Zhao}
\email{zhaojj@dlut.edu.cn}
\address{Key Laboratory of Materials Modification by Laser, Ion and Electron Beams, Ministry of Education, Dalian University of Technology, Dalian 116024, China}
\address{School of Physics, Dalian University of Technology}

\author{James R. Chelikowsky}
\email{jrc@utexas.edu}
\address{Center for Computational Materials, Oden Institute for Computational Engineering and Sciences, The University of Texas at Austin, Austin, TX 78712}
\address{ Department of Physics, The University of Texas at Austin, Austin, TX 78712}
\address{ McKetta Department of Chemical Engineering, The University of Texas at Austin, Austin, TX 78712
}

\begin{abstract}
Two-dimensional (2D) multiferroic materials are ideal systems for exploring new coupling mechanisms between different ferroic orders and producing novel quantum phenomena with potential applications. 
We employed first-principles density functional theory calculations to discover intrinsic ferroelectric and anti-ferroelectric phases of CrPSe$_3$, which show ferromagnetic order and compete with the centrosymmetric phase with an antiferromagnetic order. 
Our analysis show that the electrical dipoles of such type-I multiferroic phases come from the out-of-plane displacements of phosphorus ions due to the stereochemically active lone pairs. 
The coupling between polar and magnetic orders creates the opportunity for tunning the magnetic ground state by switching from the centrosymmetric to the ferroelectric phase using an out-of-plane electric field.
In ferroelectric and antiferroelectric phases, the combination of easy-plane anisotropy and Dzyaloshinskii-Moriya interactions (DMI) indicate they can host topological magnetic vortices like meron pairs. 
\end{abstract}

\maketitle


Multiferroic systems concomitantly host two or more ferroic orders and promise novel applications by coupling different order parameters and supporting exotic quantum states~\cite{Spaldin2019}. 
In particular, intrinsic magnetoelectric multiferroics, such as BiFeO$_3$ and YMnO$_3$, are of special interest not only because they are model systems for understanding magnetoelectric coupling, but also because they have unusual features like conductive domain walls~\cite{Ghara2021,Seidel2009,Meier_2015} and photostriction~\cite{Wei2017,Kundys2010}.
As different ferroic ordering~\cite{Gong2017,Huang2017,Jiang2021,Zhou2017,Chang2016} have recently been found in 2D materials, researchers are devoting more attention to 2D multiferroic systems~\cite{tang2019}. 
However, intrinsic bulk magnetoelectrics, let alone 2D magnetoelectrics, are difficult to find owing to the contradicting requirements for chemical conditions supporting ferroelectricity and magnetism~\cite{Hill2000}.

Recently, artificial 2D multiferroic systems have been proposed, which involve combining materials with diverse ferroic orders to generate heterostructures~\cite{Li2021}, doping magnetic ions into ferroelectrics~\cite{Yang2020}, or creating bilayer structures with customized stacking orders~\cite{Liu2020}.
Theoretically, a few intrinsic 2D magnetoelectrics have also been predicted.
For example, Zhang \etal~proposed a type-II multiferroic MXene with weak electric dipole moments induced by the helical magnetic order\cite{junjie2018}.
Qi \etal~discovered that a class of monolayer quaternary compounds has an anti-ferroelectric ground-state structure combined with long-range magnetism~\cite{Qi2018}. 
Several groups predicted that monolayer VOX$_2$ (X = Cl, Br, I) will have magnetic ordering as well as in-plane ferroelectricity~\cite{Xu2020,Tan2019}.
In monolayer VOI$_2$, the combination of iodine's strong spin-orbital coupling and breaking inversion symmetry causes large DMI~\cite{DZYALOSHINSKY1958241,Moriya1960} and  topological magnetic vortices~\cite{Xu2020}.
Meanwhile, there are still debates in the literature about the metallicity of VOI$_2$, which hinders its in-plane ferroelectricity~\cite{Ding2020}.
Experiments have yet to corroborate these pioneering predictions of intrinsic 2D multiferroics.

In this Letter, we revisit the layered Van der Waals material CrPSe$_3$ using first-principles calculations and Monte-Carlo simulations. 
Our study reveals unexpected ferroelectric and anti-ferroelectric phases which has ferromagnetic orders and compete with the centrosymmetric phase, which is antiferromagnetic. 
The coupling between magnetic ground states and polar orders sheds light on tunning magnetism using electric field.
In addition, the polar and anti-polar phases also display sizable DMI and easy-plane anisotropy, which support the formation of  topological magnetic vortices such as merons~\cite{AlfardoNov1976}. 


CrPSe$_3$ belongs to the MPX$_3$ family of metal thio- and selenophosphites, where M stands for transition metal elements and X for sulfur or selenium~\cite{Michael2017}. 
Most MPX$_3$, including CrPSe$_3$, have layered structures stabilized by Van der Waals interactions, but they appear in various stacking orders. 
To investigate the structure and magnetic order of CrPSe$_3$, we performed density functional theory (DFT) calculations based on a plane-wave pseudopotential framework~\cite{QE-2009,QE-2017}.
When comparing various exchange-correlation functionals~\cite{PBE,Grimme2010}, we found PBEsol and local-density approximation functionals provide a better overall description of the structural properties of bulk CrPSe$_3$ (as shown in Supplementary materials~\cite{supp}\cite{VANSETTEN201839,PBEsol,kulik2006}). Here all results presented were calculated with the PBEsol functional.

\begin{figure}[tb!]
    \centering
    \includegraphics[width=0.45\textwidth]{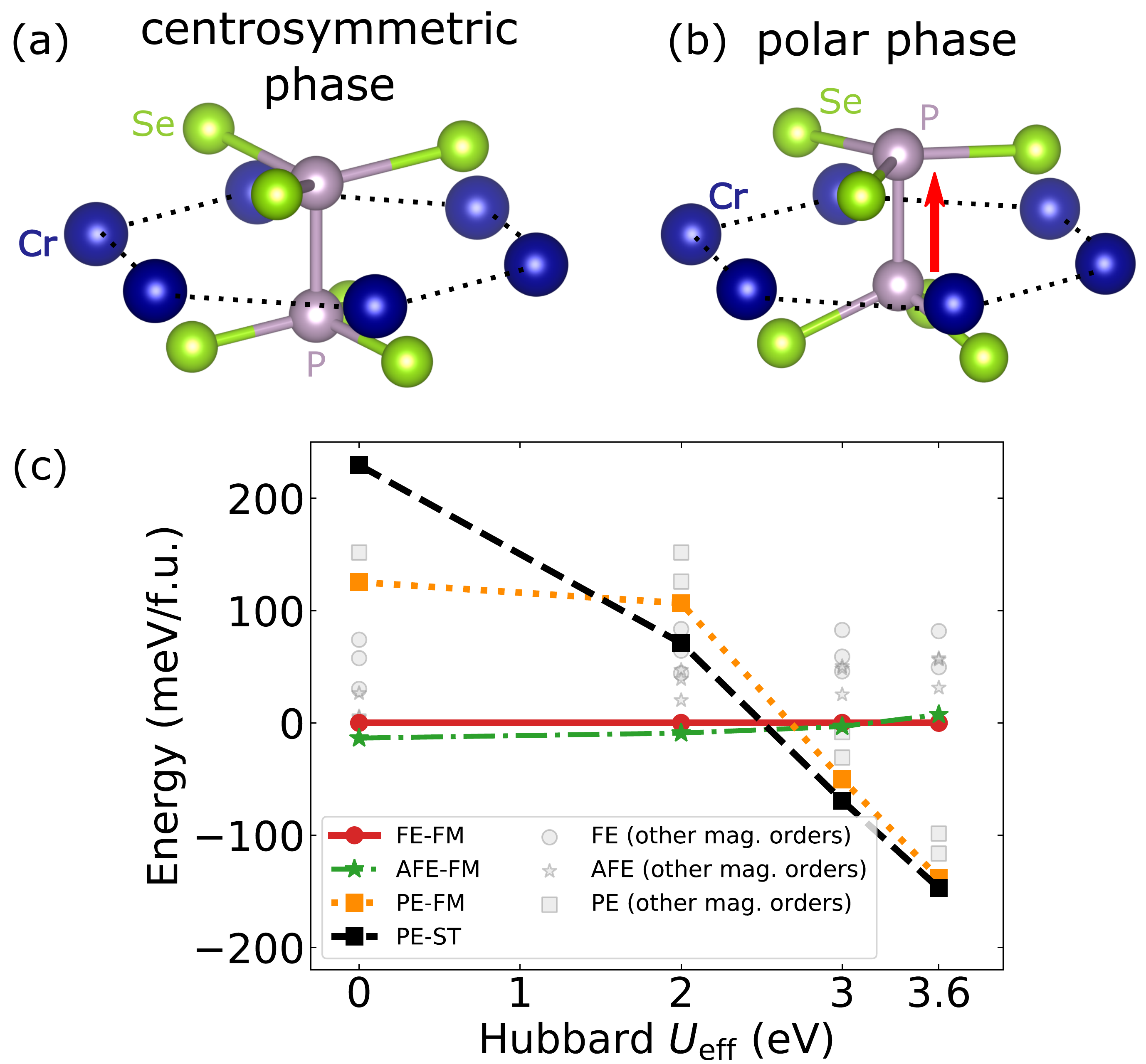}
    \caption{Atomic models for (a) the centrosymmetric phase and (b) the polar phase~\cite{VESTA}; (c)  Comparison of the total energies between different multiferroic configurations when calculated with different $U_\mathrm{eff}$.}
    \label{fig:struc_show}
\end{figure}

We carried out DFT+U calculations~\cite{matte02005} with effective parameters $U_\mathrm{eff}$ to optimize the structure and determine the energy of monolayer and bulk CrPSe$_3$~\cite{Michael2017}. 
Interestingly, in addition to the centrosymmetric phase similar to other MPX$_3$, we found an unexpected polar phase with phosphorus ions tilting along the out-of-plane direction.  The displacements of phosphorus ions break the inversion symmetry, as schematically depicted in \FIG ~\ref{fig:struc_show} (b). 
Furthermore, an anti-polar structure can be constructed based on the polar phase, in which the phosphorus atoms in neighboring cells tilt in opposing directions (see Supplementary material for a plot of the anti-polar structure).

To comprehensively compare the energies of various structures and magnetic orders of monolayer CrPSe$_3$, we examined 12 different multiferroic configurations which combine polar (FE), anti-polar (AFE), centrosymmetric (i.e., paraelectric or PE) structures with the FM and three common antiferromagnetic orders, namely, zigzag (ZZ), stripy (ST), and N\'eel (NL) orders. 
Acronyms are used to represent multiferroic phases in the following discussions.
``FE-FM", for example, denotes the phase with the polar structure and a ferromagnetic order.
As shown in \FIG~\ref{fig:struc_show} (c), the energies of 12 configurations are compared with the energy of the FE-FM phase, which is set as the reference point. 
Overall, the energies of the PE phases change with $U_\mathrm{eff}$ at a different rate compared to the AFE and FE phases. 
The difference is attributed to different electron occupations of the Cr $3d$-orbitals ({\it i.e.}, different oxidation states of Cr atoms) in polar and non-polar structures. 
For a reasonable range of $U_\mathrm{eff} \in [2.0, 3.0]$~eV of Cr~\cite{guy2022}, the FE-FM, AFE-FM, and PE-ST phases are competing phases with similar energies. 
The competition between AFE-FM, FE-FM, and PE-ST phases of CrPSe$_3$ suggest the possibility of the coexistence of these three phases. 
Similar competitions between various phases has also been observed in ferroelectrics such as In$_2$Se$_3$~\cite{Xu2020b}, CuInP$_2$Se$_6$~\cite{Dziaugys2020}, and so on~\cite{Hiroki2018,Batra2017,Qi2020}.

Among different multiferroic configurations, the AFE-FM phase has the lowest energy when $U_\mathrm{eff} < 2.5$~eV, while PE-ST is the lowest when $U_\mathrm{eff} > 2.5$~eV, as shown in \FIG~\ref{fig:struc_show} (c). 
The AFE-FM and FE-FM phases have small energy differences within 12 meV/f.u. 
As $U_\mathrm{eff}$ rises over 3.2 eV, the energy of FE-FM phase falls below that of AFE-FM phase.
Examining the magnetic orders, we find FM order remains the lowest-energy magnetic order for both the FE and AFE structures. 
When $U_\mathrm{eff} > 1.5$ eV, the PE phase has the ST magnetic order, which changes to the FM order when $U_\mathrm{eff}$ is lower than $1.5$ eV. 
When $U_\mathrm{eff}$ is set to 0,  which is an improper value for Cr, the predicted magnetic order of the PE structure is the FM order, which matches prior work~\cite{Chittari2016} utilizing the PBE+D2 method~\cite{Grimme2006}. 
We also conducted PBE+D2~\cite{Grimme2010} computations, which quantitatively agree with earlier results for these magnetic ordering~\cite{Chittari2016}, reinforcing the reproducibility of our work. 

\begin{figure}[tb!]
    \centering
    \includegraphics[width=0.49\textwidth]{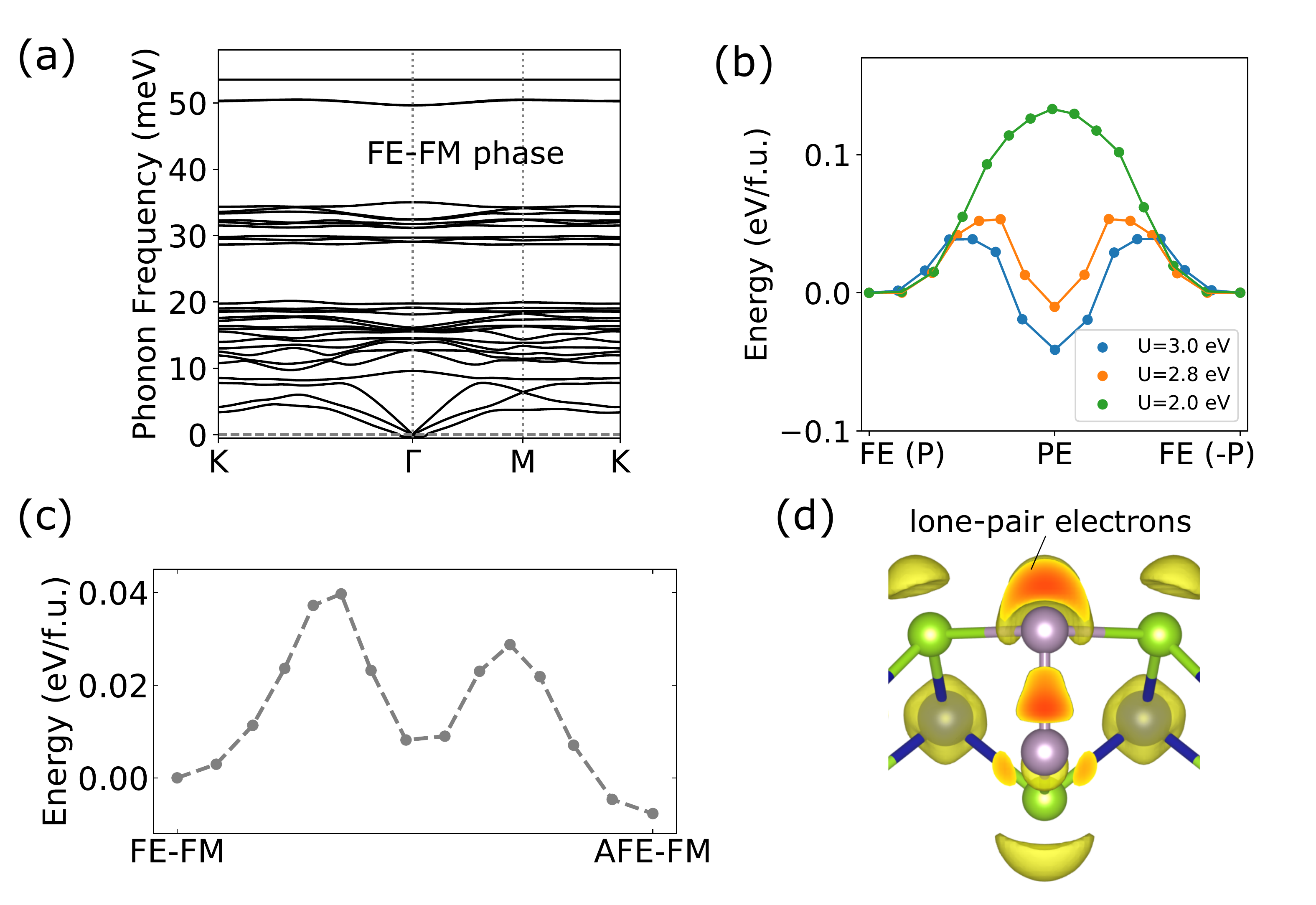}
    \caption{(a) Phonon dispersion of the FE-FM phase;  (b) Energy barriers for switching the electric polarization of the ferroelectric (i.e., polar) phase; (c) The energy profile for a structural transition from the FE-FM phase to the AFE-FM phase calculated with PBEsol+U (with $U_\mathrm{eff}=2.8$ eV); (d) Isosurfaces of electron localization function in the polar phase, showing the lone-pair electrons near phosphorus atoms.}
    \label{fig:2}
\end{figure}

Using a frozen-phonon method~\cite{phonopy}, we computed the phonon spectra of the polar (as shown in \FIG ~\ref{fig:2} (a)), centrosymmetric, and antipolar structures. 
Their phonon dispersions show no imaginary frequency modes and demonstrate dynamical stability. 
Our Berry-phase calculations~\cite{king1993} show the polar phase has an electric polarization density of 3.98~pC/m, which is larger than the predicted type-II multiferroic Hf$_2$VC$_2$F$_2$ (1.98~pC/m)~\cite{junjie2018}. 
The nudged-elastic band (NEB) approach is adopted to determine the energy barrier $E^{P\rightarrow-P}$ of collectively reversing the electric polarization direction~\cite{neb}. 
The switching energy barrier $E^{P\rightarrow-P}$ changes from $0.05$ to $0.13$ eV/f.u when $U_\mathrm{eff}$ varies from 3.0 eV to 2.0 eV. 
When the $U_\mathrm{eff}$ is set to a value greater than 2.4 eV, a local minimum similar to the PE phase appears in the middle of the transition path, as shown in \FIG~\ref{fig:2} (b). 
The energy barrier to leave the potential well around the local minimum structure, however, is reasonably low (less than 100~meV/f.u. for $U_\mathrm{eff}<3.0$ eV) to permit a switching of polarization under suitable electric field, as it is comparable to other out-of-plane 2D ferroelectrics~\cite{Ding2017,Brehm2020}. 
Notably, the interdependence between the polar and magnetic order in CrPSe$_3$ is uncommon and offers an opportunity for altering magnetic ground-states by inducing a transition from the centrosymmetric structure to the polar structure.

Similarly, we estimated the energy barrier $E^{\mathrm{FE}\rightarrow\mathrm{AFE}}$ between the FE-FM and AFE-FM phases. 
The energy profile of transitioning from the FE-FM phase to the AFE-FM phase is shown in \FIG~\ref{fig:2} (c). 
There are two energy barriers along the transition path. 
The crystal structure transforms into an intermediate configuration close to the PE phase after passing through the first energy barrier. The intermediate configuration corresponds to the dip on energy curve of the switching process, as shown in \FIG~\ref{fig:2} (c).
After overcoming the second barrier, the structure turns into the AFE structure. 
Overall, the heights of these two barriers are on the order of 40~meV/f.u., which is in the same order-of-magnitude as the energy barrier $E^{P\rightarrow-P}$ of switching the polarization of the FE phase. 
Although the FE-FM phase is metastable, the energy ordering of the FE-FM and AFE-FM phases can be reversed with moderate electric fields or strain (see the supplementary materials), suggesting the possibility of inducing a phase transition from the AFE-FM phase to the FE-FM phase using electric field or strain. 



For bulk CrPSe$_3$ phases, we considered additional interlayer antiferroelectric and interlayer antiferromagnetic orders (see Supplementary Material for details of these two orders), and constructed 20 different multiferroic configurations.  
The dependence of ground-state phases on $U_\mathrm{eff}$ is similar to the situation of monolayer CrPSe$_3$. 
As the Hubbard $U_\mathrm{eff}$ parameter changes, the lowest-energy configuration is the PE-ST phase when $U_\mathrm{eff}$ parameter is between around 2.6 to 3.6 eV, but changes to the AFE-FM phase when $U_\mathrm{eff}$ drops to below 2.6 eV. 
With $U_\mathrm{eff}$ larger than 2.0~eV, our calculation results show that the ground-state magnetic order of the centrosymmetric phase is anti-ferromagnetic ST ordering. 
Recently, a theoretical study shows that monolayer and bulk PE phase of CrPSe$_3$ have NL order, while the ST order has an energy slightly higher than the NL order~\cite{Xu_2021}. 
The discrepancy between this work and our study likely originates from different types of pseudopotentials (see Supplementary materials for more details). 
Experiments show that bulk CrPSe$_3$ has an anti-ferromagnetic order~\cite{Rui2017}, while its detailed anti-ferromagnetic spin arrangement is not determined. 

Because phosphorous contributes polarization and chromium contributes magnetism, the coexistence of polar and magnetic order does not violate the $d^0$-rule~\cite{Hill2000} and the FE-FM and AFE-FM phases are type-I multiferroic phases. 
The out-of-plane displacements of phosphorus ions in polar and anti-polar structures are related to the oxidation states of Cr ions. 
In the centrosymmetric phase, the oxidation state of chromium ions are Cr$^{2+}$, similar as the transition metal ions $M^{2+}$ in most MPSe$_3$ systems~\cite{Michael2017}. 
However, in the anti-polar and polar phases, it becomes Cr$^{3+}$. 
Compared to the centrosymmetric structure with Cr$^{2+}$, a pair of Cr$^{3+}$ ions in the polar structures give up two electrons, generating a lone pair located nearby phosphorus atoms. 
As a result, analogous to BiFeO$_3$~\cite{Spaldin2019}, lone-pair electrons result in an asymmetric charge distribution that pushes the phosphorus atoms out of plane.
The isosurface map of the electron-localization function (ELF) visualizes the position of lone pairs and the consequent structural change, as illustrated in \FIG ~\ref{fig:2} (d). 


With the spin-orbit coupling effects contributed by Se atoms and the breaking local inversion symmetry, the AFE-FM and FE-FM phases can potentially show significant DMI, which can lead to topological spin textures. 
Here an extended Heisenberg model describing the magnetic interactions in a 2D spin-lattice is considered:
\begin{eqnarray*}
H & = & \sum_{ <i,j> } J_1 \mathbf{n}_i \cdot \mathbf{n}_j + \sum_{ \ll i,j \gg } J_2 \mathbf{n}_i \cdot \mathbf{n}_j + \\
& & \sum_{ <i,j> } \mathbf{D}_1 \cdot (\mathbf{n}_i \times \mathbf{n}_j) + \sum_{i}\sum_{\alpha=x,y,z} A_{\alpha} n^2_{i\alpha}
\end{eqnarray*}
where $\mathbf{n}_i$ are unit vectors; $<>$ and $\ll\gg$ stand for nearest and next-nearest neighbors, respectively; $J_{1,2}$ are the symmetric exchange coupling and $\mathbf{D}_{1}$ is the DMI coupling; $A_\alpha$ describes the single-ion magnetic anisotropy. 
The exchange parameters are calculated using a four-states mapping method for FE-FM and AFE-FM phases~\cite{xiang2013}. 
\TAB ~\ref{tab:magnetic_parameters} contains the magnetic parameters of the FE-FM phase. Details on the magnetic parameters of the AFE-FM phase are presented in the Supplementary Material.
Both FE-FM and AFE-FM phase have easy-plane magnetic anisotropy and large nearest-neighbor DMI interaction $\mathbf{D}_1$.
The FE-FM phase retains the $C_3$ symmetry of the PE phases and its easy-plane anisotropy is close to the XY model.  According to the Mermin-Wagner theorem~\cite{mermin_wagner}, the FE phase does not allow long-range FM order in a perfect infinite-size crystal. 
In a finite-size sample of the FE-FM phase, however, weak ferromagnetism can arise. In its easy plane, the AFE-FM phase has an easy axis along $y$-direction, which violates the continuous $O(2)$ symmetry and allows for long-range ferromagnetic order. 

\begin{table}[htb]
\centering
\caption{Parameters (in meV) for the magnetic Hamiltonian of the FE-FM phase calculated with PBEsol+U ($U_\mathrm{eff}$ = 2.8 eV).}
\label{tab:magnetic_parameters}
\setlength{\tabcolsep}{10pt}
\begin{tabular}{@{}ccccc@{}}
\toprule
$J_1$  & $J_2$  &   $A_x$ & $A_y$ &$\mathbf{D}_1$      \\ \midrule
$-21.7$ & $-0.75$  & -0.27 & -0.27 &$[0.3,0.0,-1.9]$ \\ \bottomrule
\end{tabular}%
\end{table}

\begin{figure}[htb!]
    \centering
    \includegraphics[width=0.45\textwidth]{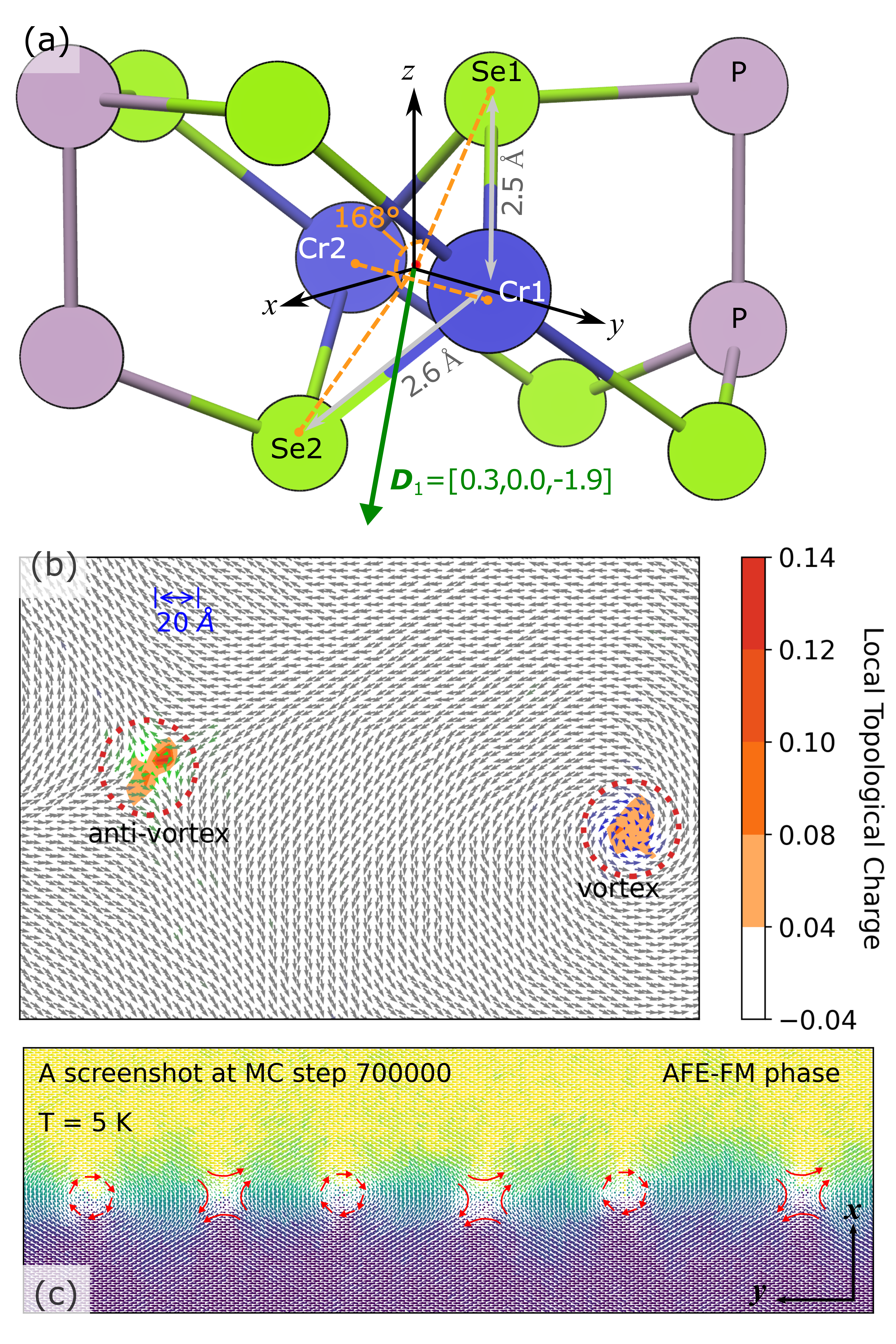}
    \caption{(a) A schematic plot showing the nearby atoms and the direction of DMI vector $\mathbf{D}_1$ between nearest-neighbour Cr ions;
    (b) A snapshot of a Monte-Carlo simulation for the FE-FM phase ($T=2$ K) which shows a pair of antimerons (topological charge of $\frac{1}{2}$) with winding numbers -1 and 1 respectively. The local topological charge density is plotted with contours on the background. Spins are colored in green or blue according to out-of-plane components. (c) A screenshot for six long-lived merons of the AFE-FM phase. Spins are colored according to the direction of their in-plane components.}
    \label{fig:merons}
\end{figure}

Because the electric polarization for the FE-FM phase is out-of-plane, one could expect the DMI vector to be in-plane.
Notably, our calculations reveal that the DMI vector has a significant out-of-plane component.
To understand this, we consider two adjacent ions Cr1 and Cr2, as well as the middle point $M$ of the line $\overline{\mathrm{Cr1Cr2}}$ connecting them. \FIG~\ref{fig:merons} (a) depict the local structure of two nearest-neighbour Cr ions. The FE-FM structure has a mirror plane passing through $M$ and is perpendicular to $\overline{\mathrm{Cr1Cr2}}$. According to Moriya's rules~\cite{Moriya1960}, the DMI vector $\mathbf{D}_1$ should also be perpendicular to $\overline{\mathrm{Cr1Cr2}}$. Our computational results agree with these rules. 
The distortion of Se ions, which mediates the indirect exchange contacts between nearby Cr ions, accounts for the large out-of-plane component of $\mathbf{D}_1$.
In the PE phase, atoms Cr1, Cr2, Se1, and Se2 are coplanar and form a parallelogram in the PE phase. In contrast, in the FE phase, the displacements of selenium ions make the dihedral angle $\angle \mathrm{Se}_1\mathrm{Cr}_1 \mathrm{Cr}_2\mathrm{Se}_2 = 168^\circ$ and bond lengths $d_{\mathrm{Cr}_1\mathrm{Se}_1} \neq d_{\mathrm{Cr}_1\mathrm{Se}_2}$, shown in \FIG~\ref{fig:merons} (a).
This local environment breaks the inversion symmetry and results in large out-of-plane components of $\mathbf{D}_1$.
The modest dipole field induced by the shifted phosphorous ions, on the other hand, creates a significantly smaller in-plane DMI component.

The synergy of out-of-plane DMI components and in-plane magnetic anisotropy favors the formation of meron and anti-meron pairs~\cite{Yu2018,Xu2020} in both FE-FM and AFE-FM phases.. 
Monte-Carlo (MC) simulations~\cite{Spirit} of the magnetic Hamiltonian were conducted for the FE-FM and AFE-FM phases. 
We initialized the spin-lattice with random spin vectors and examine its evolution in MC simulations. 
Typically, after a few thousand MC steps, small in-plane ferromagnetic domains emerge and topologically non-trivial spin vortices and anti-vortices form on the domain boundaries. 
Such topologically non-trivial magnetic excitations have a topological charge of one half or negative one half, suggesting they are anti-merons or merons. 
Notably, merons appear even under $T = 30$~K, which is higher than the allowed temperature range for merons to emerge in CrCl$_3$~\cite{Lu2020} and VOI$_2$~\cite{Xu2020}. 
Furthermore, the formation of merons is robust to the variation of DMI parameters and Heisenberg exchange parameters. 
Merons and antimerons are still observed at around $T=5$ K even when the magnitude of DMI vector decreases by 50\%. 
In MC simulations, merons with winding numbers $w=1$ and $w=-1$ , which are known as vortex and anti-vortex merons, emerge in pairs, similar as in prior investigations~\cite{Augustin2021,Gao2019}. 
\FIG~\ref{fig:merons} (b) is a typical snapshot of MC simulations for the FE-FM phase, showing real-space spin textures and the distribution of topological charge in the spin-lattice. 
Clearly, the topological charge density is concentrated near merons or antimerons.
As the MC simulation progresses, vortex (anti)meron tend to attract nearby antivortex (anti)merons. 
Such annihilation phenomenon is also observed in Landau-Lifshitz-Gilbert simulations for CrCl$_3$~\cite{Augustin2021,Lu2020} and kagome magnets~\cite{Pereiro2014}. 
In the FE-FM phase, merons and anti-merons will all be annihilated after a long enough MC simulation (typically after $5\times10^5$ steps).
In the AFE-FM phase, unusual long-lived vortex and anti-vortex meron pairs appears.
Even after $2\times10^6$ MC steps, these topological excitation pairs stabilize on the boundaries of adjacent ferromagnetic domains and exhibit no tendency to annihilate each other. 
Three long-lived vortex-antivortex pairs are shown in \FIG~\ref{fig:merons} (c). 
Such long-lived merons are typically located on the domain boundaries parallel to the $y$ or $x$-axis.

In all, we comprehensively investigated the structures and magnetic orders of different phases in bulk and monolayer CrPSe$_3$.
We identify two multiferroic phases of CrPSe$_3$, namely a ferroelectric and an anti-ferroelectric phase with ferromagnetic orders, while the experimentally studied centrosymmetric phases is antiferromagnetic. 
The multiferroic phases are both dynamically stable and compete with the centrosymmetric phase.
The metastable ferroelectric phase carries an out-of-plane electric polarization and is separated from the centrosymmetric and anti-ferroelectric phase by surmountable energy barriers.
The coupling between the polar order and magnetic ground state provide new opportunities for tunning magnetism through electric field. 
Moreover, with significant out-of-plane DMI interaction and in-plane anisotropy, the ferroelectric and anti-ferroelectric phases can potentially host magnetic topological excitations including merons and antimerons under suitable conditions. 
Our work suggests CrPSe$_3$ is a promising material for exploring intrinsic low-dimensional multiferroicity and topological magnetic excitations.

\section{Acknowledgement}
WG and JRC acknowledge support from a subaward
from the Center for Computational Study of Excited-State
Phenomena in Energy Materials at the Lawrence Berkeley
National Laboratory, which is funded by the U.S. Department
of Energy, Office of Science, Basic Energy Sciences, Materials
Sciences and Engineering Division under Contract No. DEAC02-05CH11231,
as part of the Computational Materials
Sciences Program. 
WG acknowledge support from the Fundamental Research Funds for the Central Universities, grant DUT21RC(3)033.
WG and JZ acknowledge the support by grant 12104080 and 91961204 from National Science Foundation of China. 
Computational resources are provided by Shanghai supercomputer center, national energy research scientific computing center (NERSC), and the Texas advanced computing center (TACC).


%

\begin{figure}[htpb]
    \centering
    \includegraphics[page=1,width=1.0\textwidth]{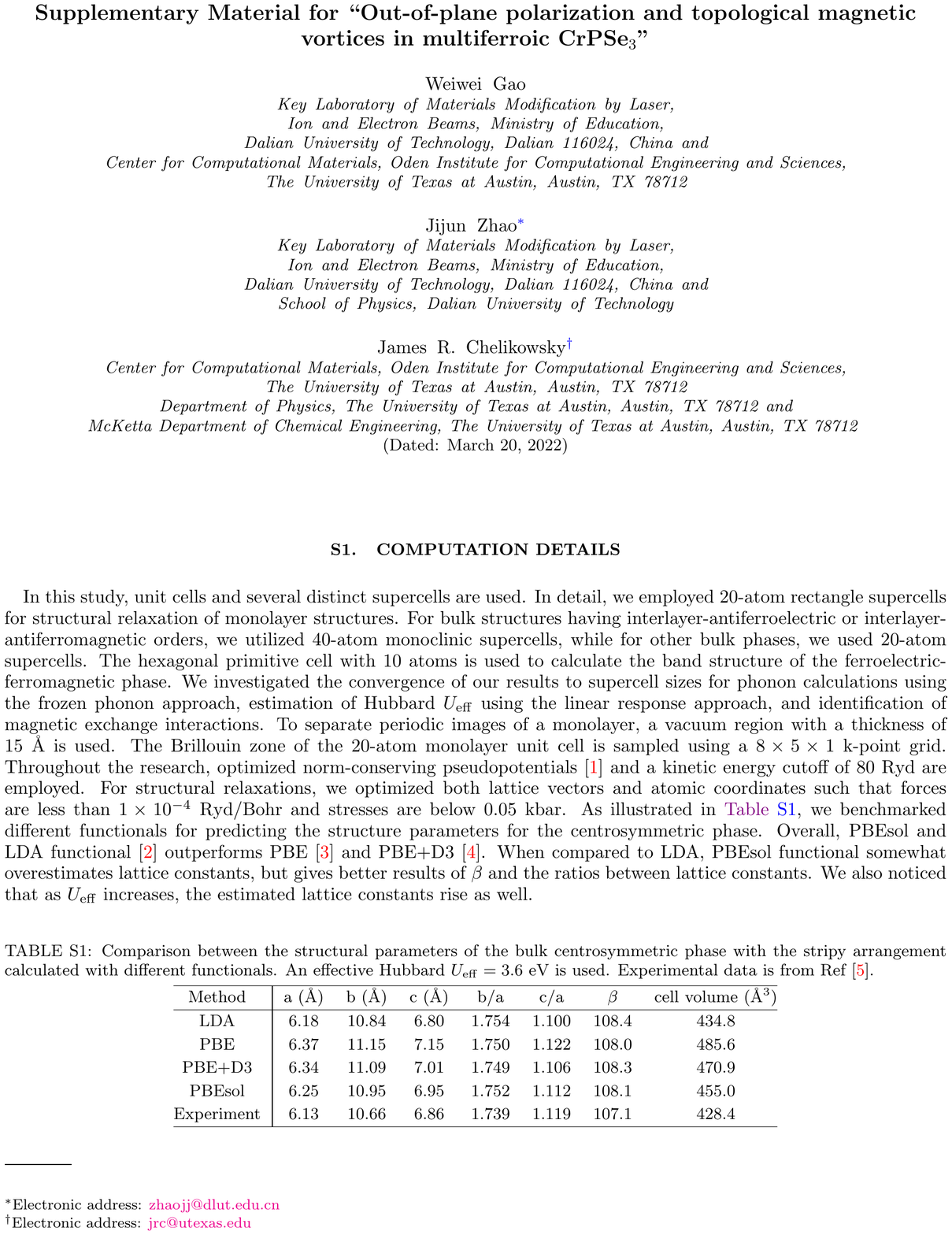}
\end{figure}

\begin{figure}[htpb]
    \centering
    \includegraphics[page=2,width=1.0\textwidth]{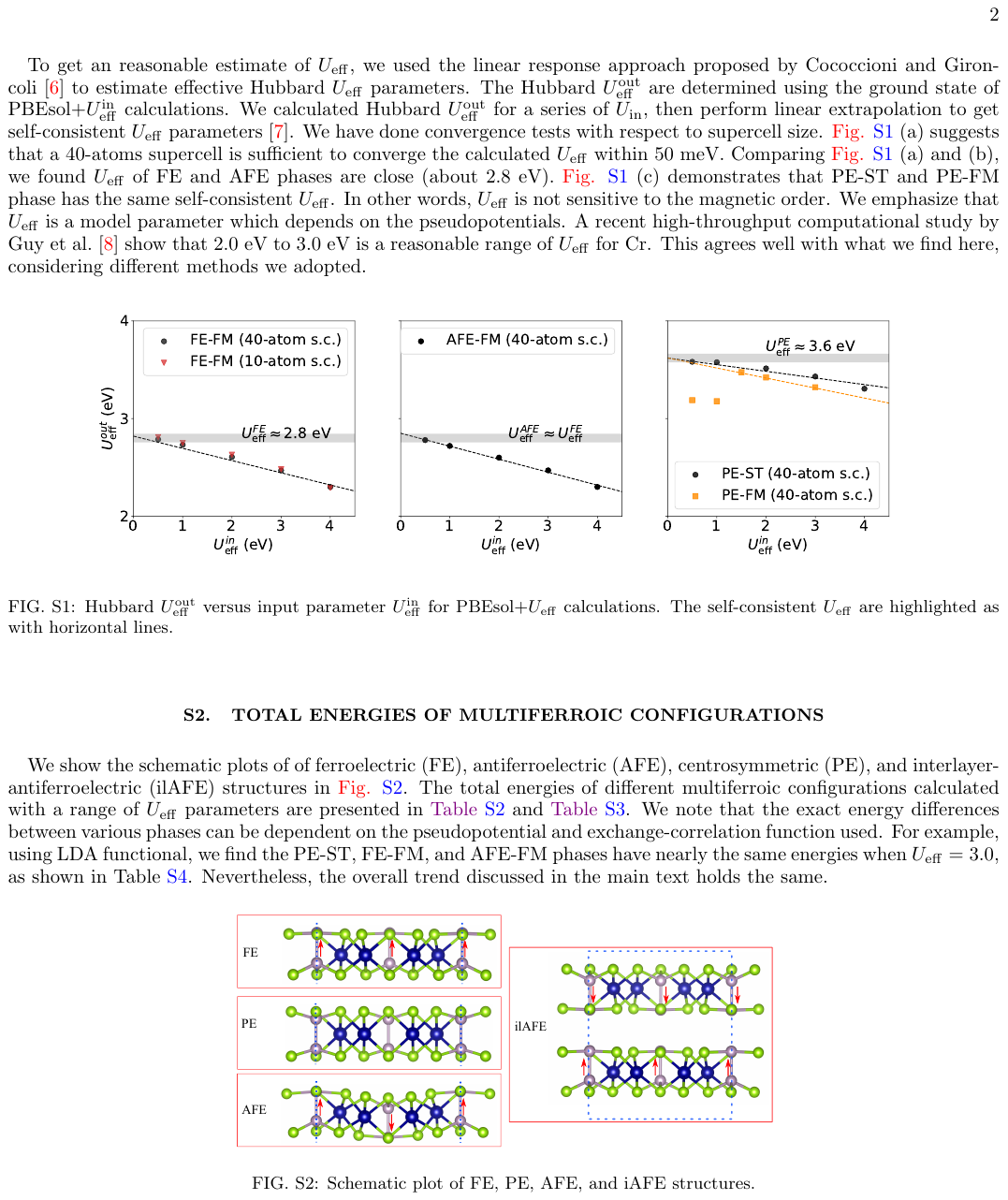}
\end{figure}

\begin{figure}[htpb]
    \centering
    \includegraphics[page=3,width=1.0\textwidth]{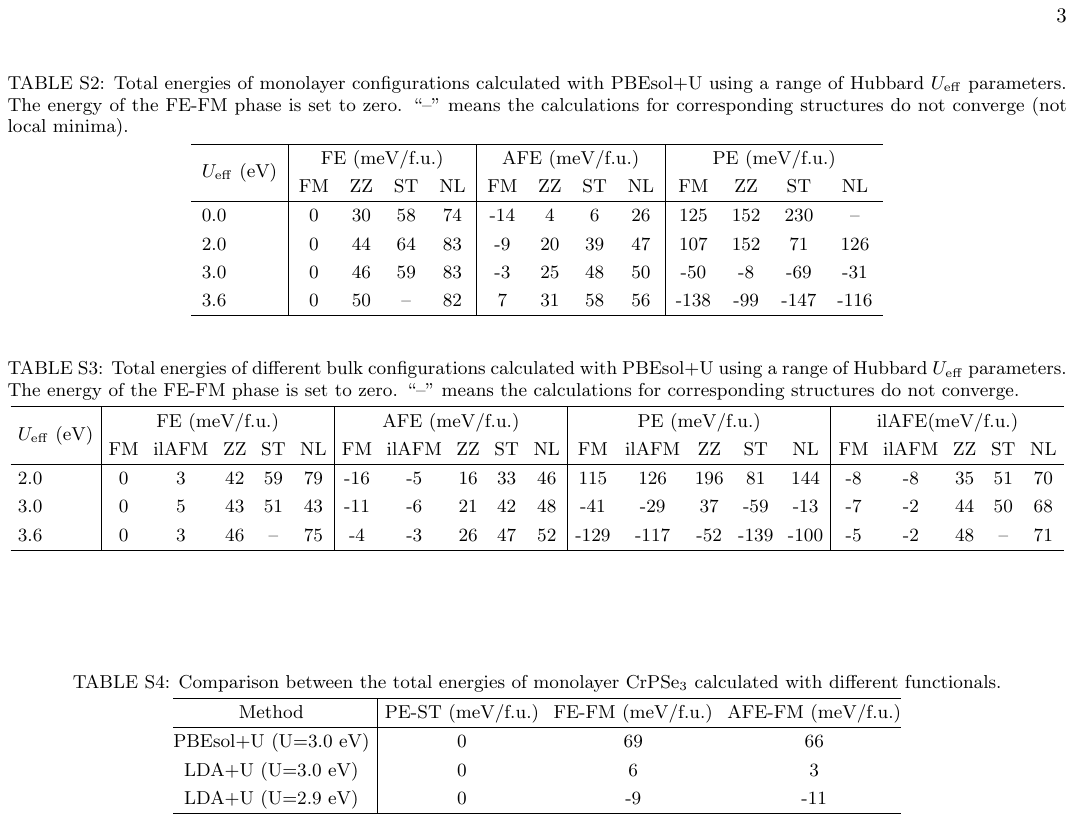}
\end{figure}

\begin{figure}[htpb]
    \centering
    \includegraphics[page=4,width=1.0\textwidth]{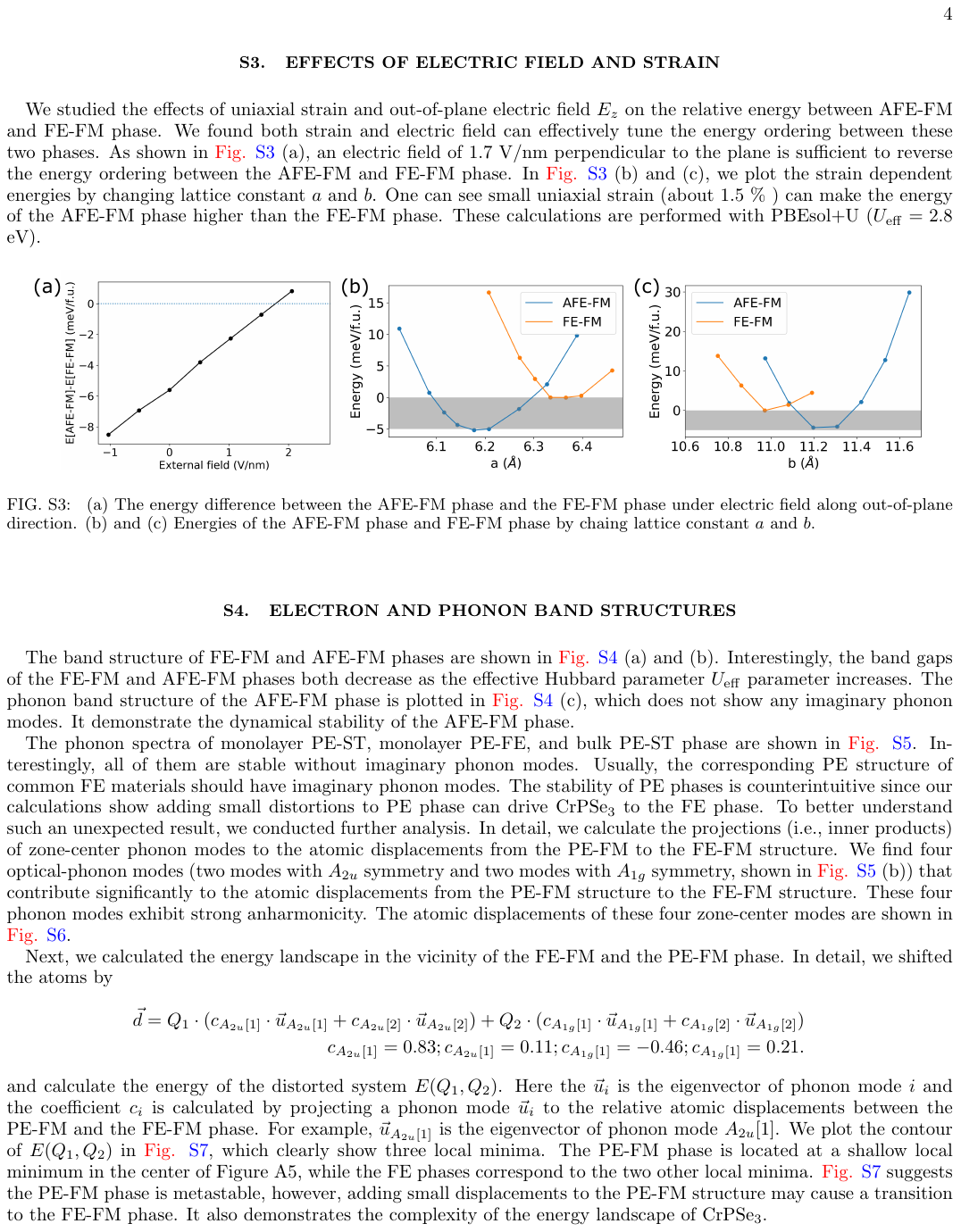}
\end{figure}

\begin{figure}[htpb]
    \centering
    \includegraphics[page=5,width=1.0\textwidth]{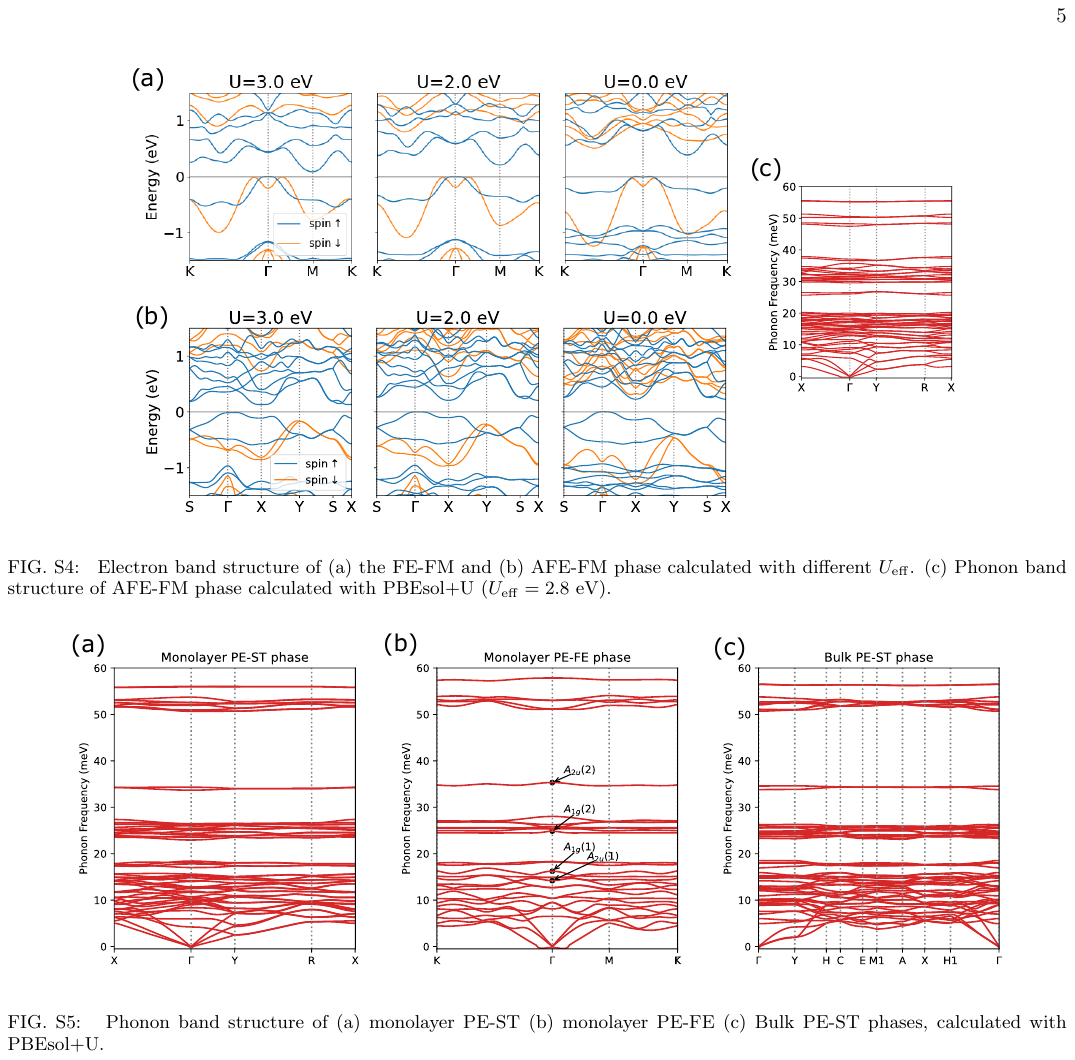}
\end{figure}

\begin{figure}[htpb]
    \centering
    \includegraphics[page=6,width=1.0\textwidth]{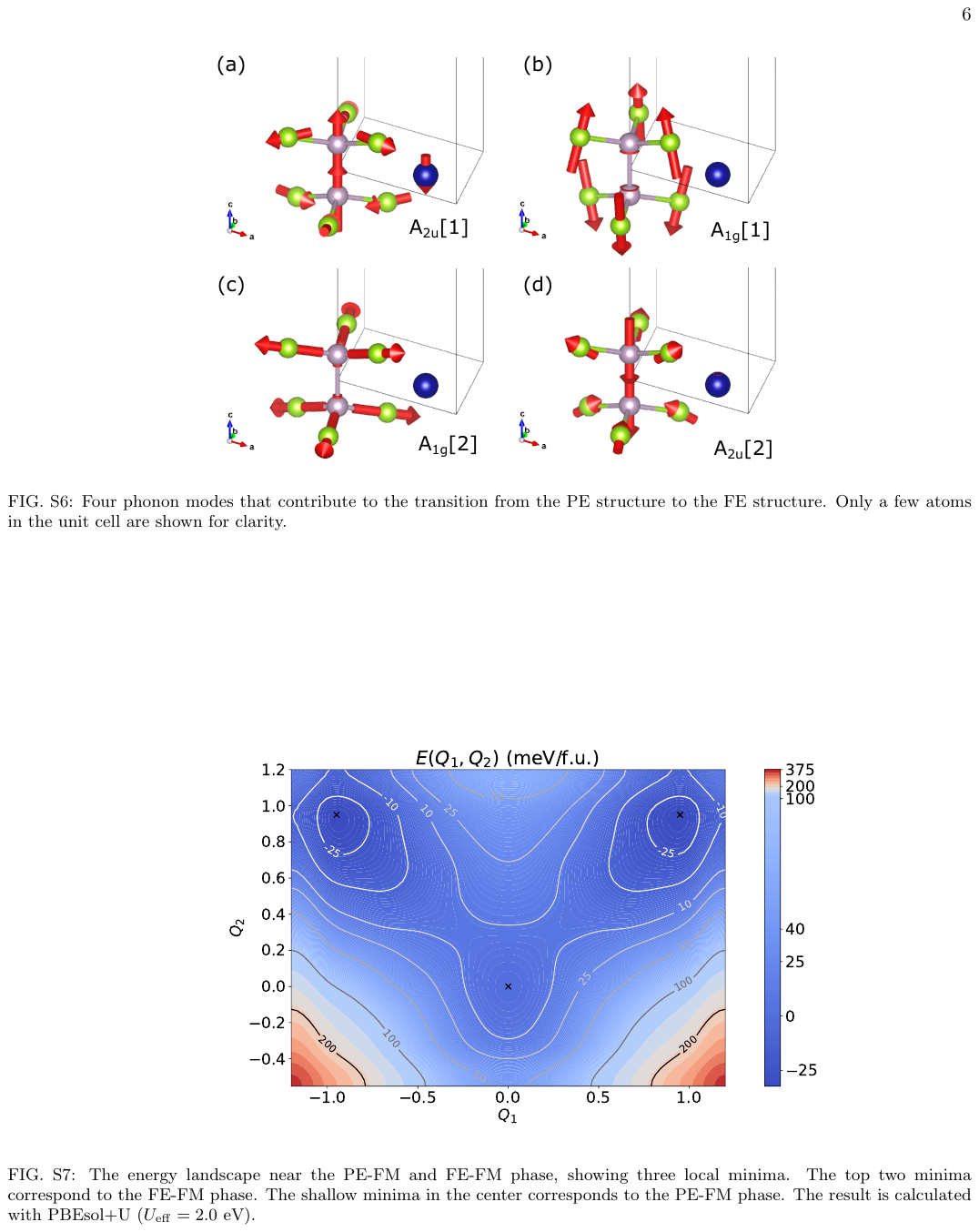}
\end{figure}

\begin{figure}[htpb]
    \centering
    \includegraphics[page=7,width=1.0\textwidth]{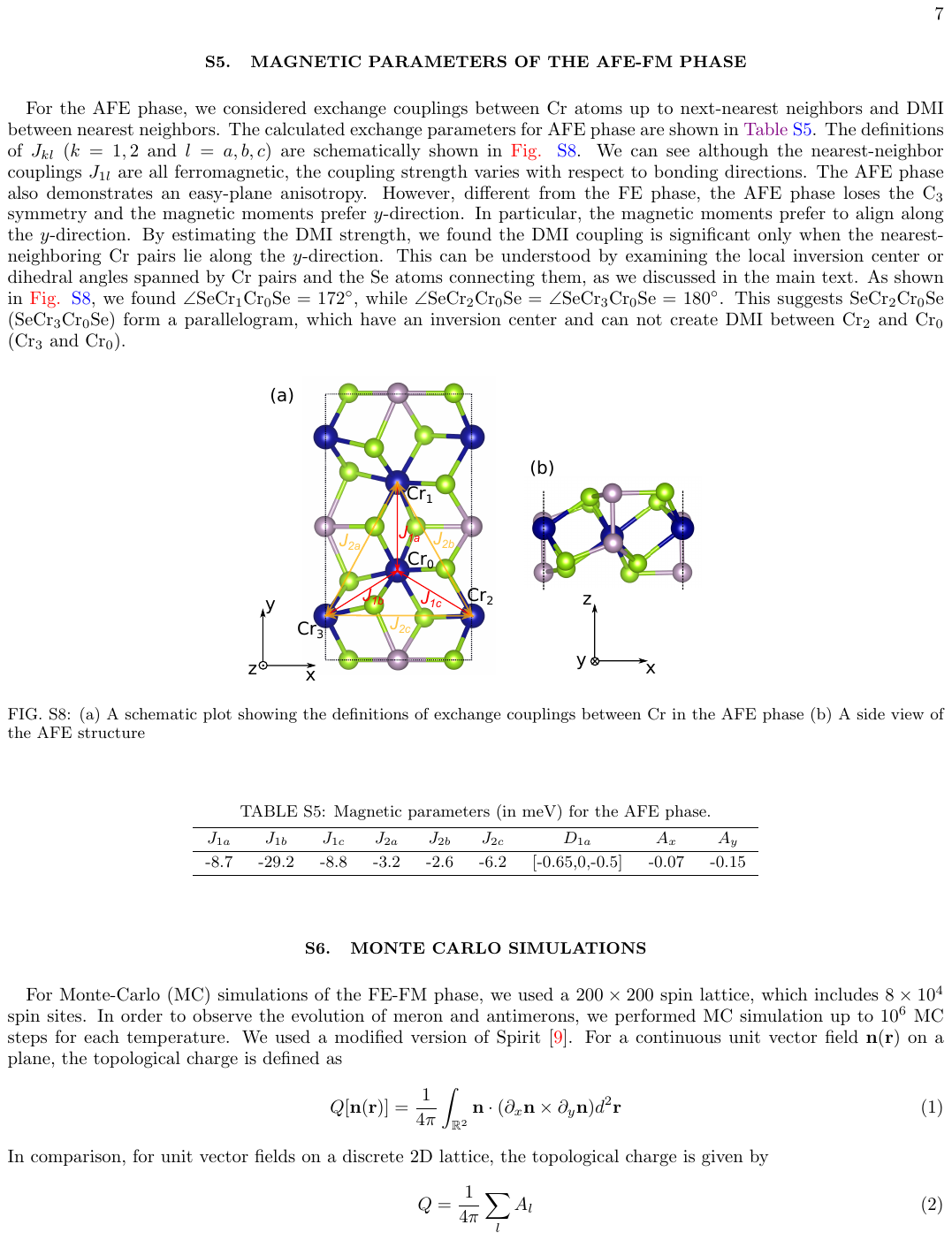}
\end{figure}

\begin{figure}[htpb]
    \centering
    \includegraphics[page=8,width=1.0\textwidth]{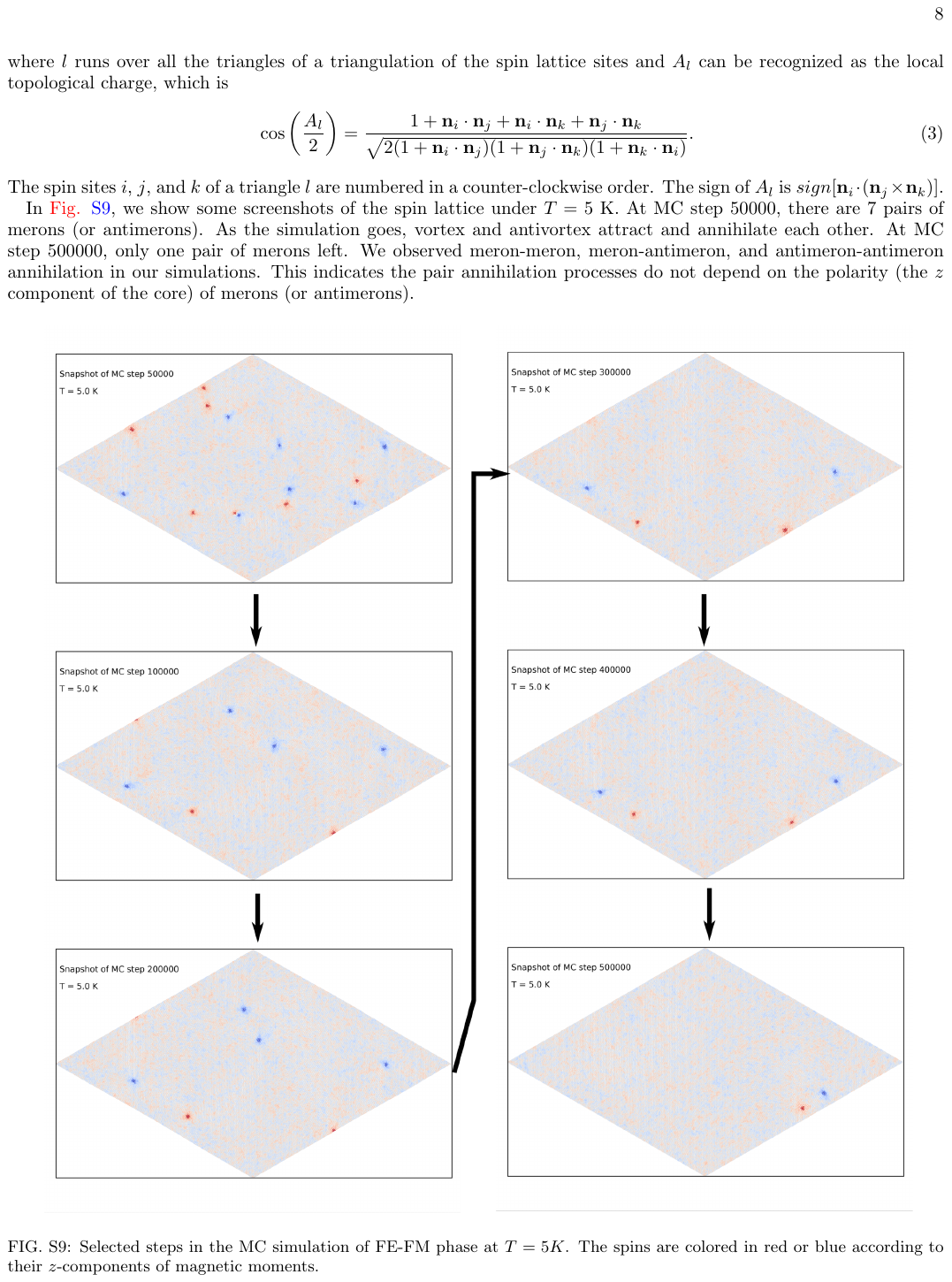}
\end{figure}

\begin{figure}[htpb]
    \centering
    \includegraphics[page=9,width=1.0\textwidth]{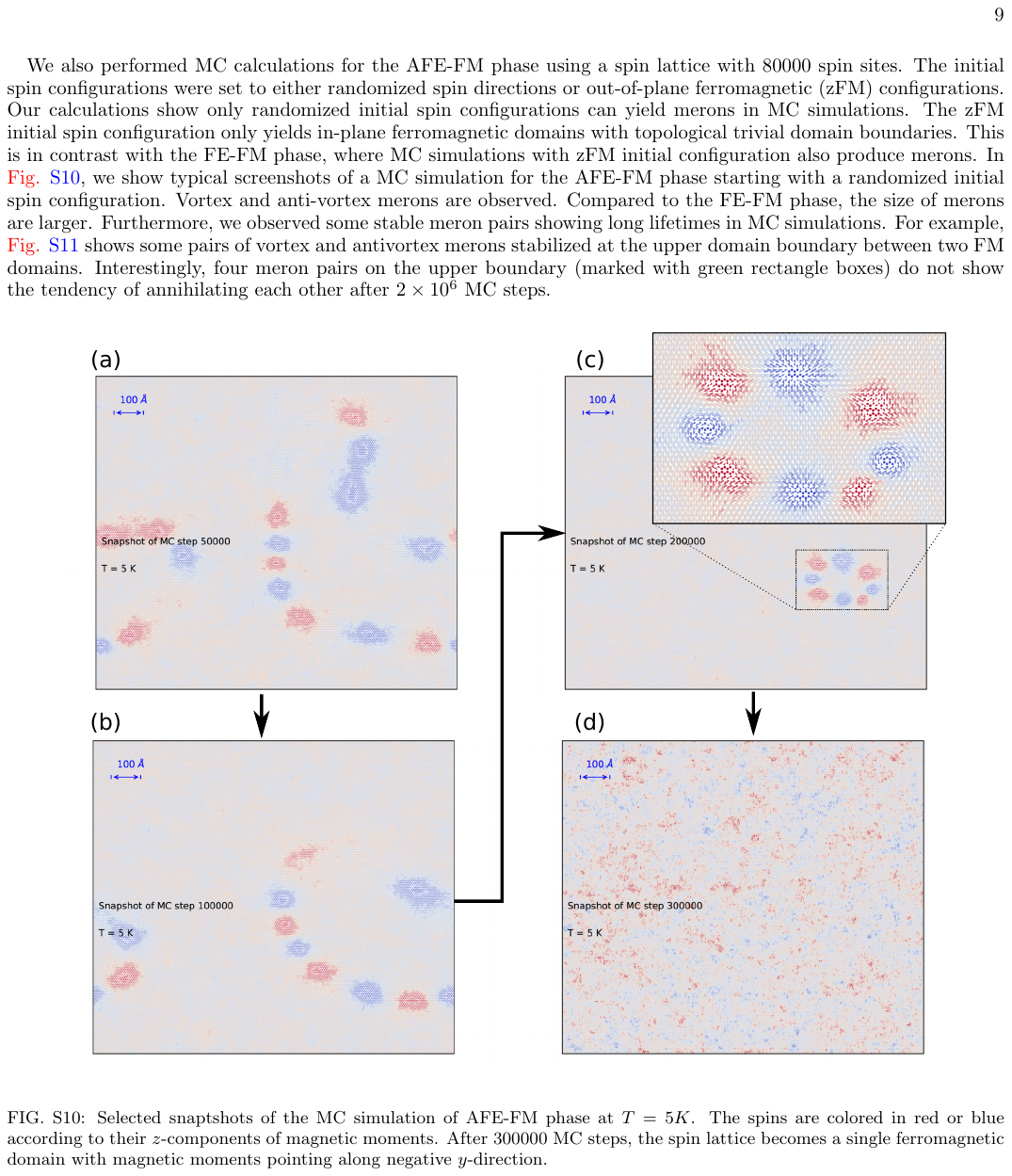}
\end{figure}

\begin{figure}[htpb]
    \centering
    \includegraphics[page=10,width=1.0\textwidth]{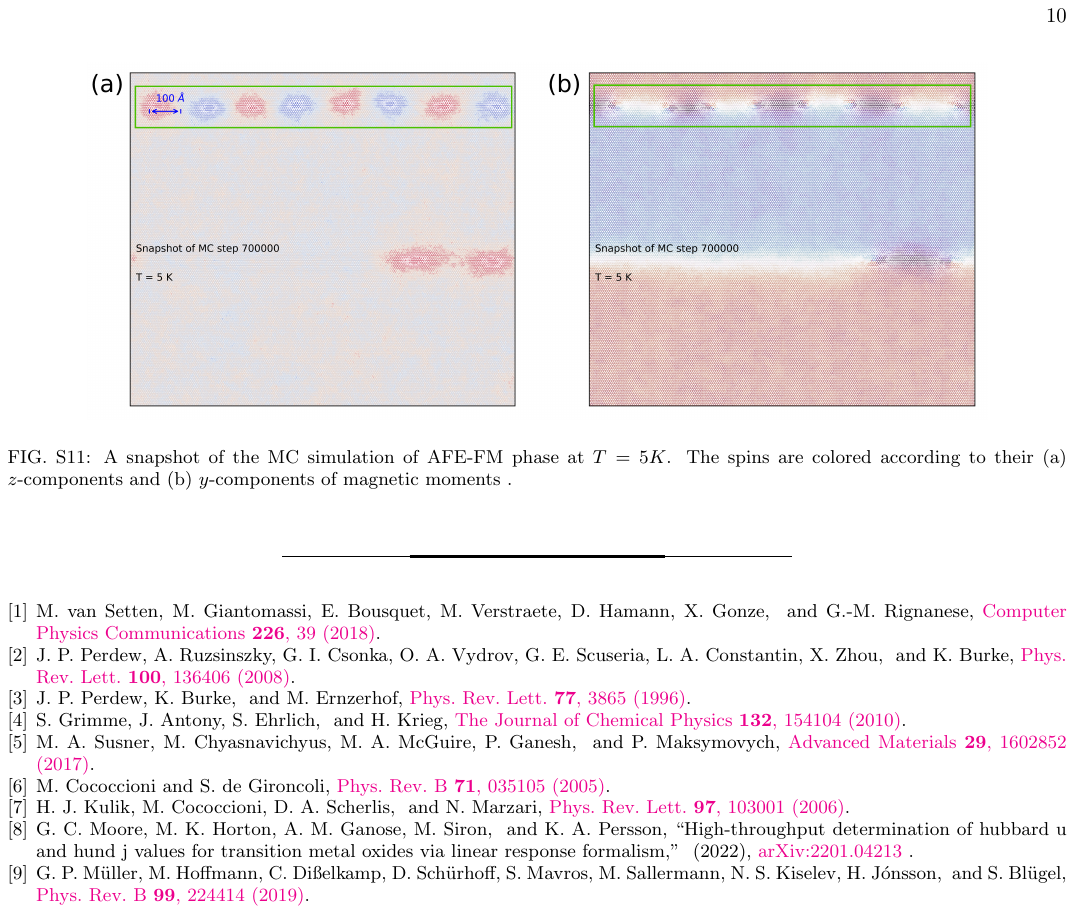}
\end{figure}

\end{document}